\documentclass[sigconf,screen]{acmart}

\AtBeginDocument{%
  \providecommand\BibTeX{{%
    \normalfont B\kern-0.5em{\scshape i\kern-0.25em b}\kern-0.8em\TeX}}}

\copyrightyear{2022}
\acmYear{2022}
\setcopyright{acmlicensed}
\acmConference[ICPC '22]{30th International Conference on Program Comprehension}{May 16--17, 2022}{Virtual Event, USA}
\acmBooktitle{30th International Conference on Program Comprehension (ICPC '22), May 16--17, 2022, Virtual Event, USA}
\acmPrice{15.00}
\acmDOI{10.1145/3524610.3527919}
\acmISBN{978-1-4503-9298-3/22/05}

\setcitestyle{numbers,sort&compress}
\usepackage{amsmath,amsfonts}
\usepackage{graphicx}
\usepackage{textcomp}
\usepackage{xcolor}
\usepackage{color}
\usepackage{bm}
\usepackage{xspace}
\usepackage{listings}
\usepackage{booktabs}
\usepackage{url}
\usepackage[skins]{tcolorbox}
\usepackage{multirow}
\usepackage{algorithm}  
\usepackage{algpseudocode}
\usepackage{makecell}
\usepackage{enumitem}
\usepackage{xcolor}
\usepackage{fontawesome}
\usepackage{tikz}
\usepackage[]{enumitem}
\usepackage{tabularx}
\usepackage{caption}
\usepackage{adjustbox}
\usepackage{threeparttable}
\usepackage{subcaption}
\usepackage{balance}
\usepackage{setspace}

\definecolor{codeblack}{RGB}{0,0,0}
\definecolor{codedeepgrey}{RGB}{102,102,102}
\definecolor{codegrey}{RGB}{208,208,208}
\definecolor{backcolour}{rgb}{0.95,0.95,0.95}

\lstdefinestyle{mystyle}{
    backgroundcolor=\color{backcolour},   
    commentstyle=\color{codegreen},
    keywordstyle=\color{magenta},
    numberstyle=\tiny\color{codegray},
    stringstyle=\color{codepurple},
    basicstyle=\ttfamily\scriptsize,
    breakatwhitespace=false,         
    breaklines=true,                 
    captionpos=b,                    
    keepspaces=true,                 
    numbers=left,                    
    numbersep=5pt,                  
    showspaces=false,                
    showstringspaces=false,
    showtabs=false,                  
    tabsize=2
}
\graphicspath{{img/}}

\newcommand{\Code}[1]{\begin{small}\texttt{#1}\end{small}}
\newcommand{\ColorRule}[3][black]{\textcolor{#1}{\rule{#2}{#3}}}

\tcbset{
  my box/.style={
    enhanced,
    colframe=#1!80,
    colback=#1!10,
    attach boxed title to top left={xshift=0.2cm, yshift=-0.2cm},
    boxed title style={
      colback=#1!80,
      outer arc=0pt,
      arc=0pt,
      top=0pt,
      bottom=0pt,
    },
  },
}
\newtcolorbox{result-rq}[1]{
  my box=black,
  title=#1,
  boxrule=1.2pt,top=6pt,bottom=3.5pt,left=6pt,right=6pt
}

\newcommand{\dataset}{1,731\xspace}
\newcommand{\finaldataset}{909\xspace}
\newcommand{\falsedata}{822\xspace}

\newcommand{\statement}{491, 28.37\%\xspace}
\newcommand{\noninfo}{137, 7.91\%\xspace}
\newcommand{\irr}{121, 6.99\%\xspace}
\newcommand{\unreach}{42, 2.43\%\xspace}
\newcommand{\noneng}{17, 0.98\%\xspace}
\newcommand{\mistake}{14, 0.81\%\xspace}

\newcommand{\con}{251, 25.61\%\xspace}
\newcommand{\connum}{251\xspace}
\newcommand{\conratio}{25.61\%\xspace}

\newcommand{\present}{81, 8.27\%\xspace}
\newcommand{\presentnum}{81\xspace}
\newcommand{\presentratio}{8.27\%\xspace}

\newcommand{\acc}{173, 17.65\%\xspace}
\newcommand{\accnum}{173\xspace}
\newcommand{\accratio}{17.65\%\xspace}

\newcommand{\product}{475, 48.47\%\xspace}
\newcommand{\productnum}{475\xspace}
\newcommand{\productratio}{48.47\%\xspace}

\newcommand{\ie}{{i.e.},\xspace}

\newcommand{\eg}{{e.g.},\xspace}

\newlength{\fsize}
\newcommand*{\itembullet}{%
    \begin{tikzpicture}[x=\fsize,y=\fsize]
        \filldraw[draw=black, fill=white] (0,0) -- (-0.15, 0.3) -- (0.38, 0) -- (0, 0);
        \filldraw[draw=black, fill=black] (0,0) -- (-0.15, -0.3) -- (0.38, 0) -- (0, 0);
    \end{tikzpicture}%
}

\newcommand{\myfancylabel}{\begin{tikzpicture}[every node/.style={rotate=45}]%
\node[fill,inner sep=0pt,minimum size=0.5ex] at (0ex,0.5ex) {};%
\node[fill,inner sep=0pt,minimum size=0.5ex] at (0ex,-0.5ex) {};%
\node[fill,inner sep=0pt,minimum size=0.5ex] at (0.5ex,0ex) {};%
\node[fill,inner sep=0pt,minimum size=0.5ex] at (-0.5ex,0ex) {};%
\end{tikzpicture}}

\begin{document}

\title{Demystifying Software Release Note Issues on GitHub}

\author{Jianyu Wu, Hao He, Wenxin Xiao, Kai Gao, Minghui Zhou}
\authornote{Minghui Zhou is the corresponding author.}
\affiliation{
  \institution{School of Computer Science and School of Software \& Microelectronics, Peking University, Beijing, China}
  \country{Key Laboratory of High Confidence Software Technologies, Ministry of Education, Beijing, China}
}
\email{{wujianyu, heh, gaokai19, zhmh}@pku.edu.cn,  wenxin.xiao@stu.pku.edu.cn}

\begin{abstract}
Release notes (RNs) summarize main changes between two consecutive software versions and serve as a central source of information when users upgrade software.
While producing high quality RNs can be hard and poses a variety of challenges to developers, a comprehensive empirical understanding of these challenges is still lacking.
In this paper, we bridge this knowledge gap by manually analyzing \dataset latest GitHub issues to build a comprehensive taxonomy of RN issues with four dimensions: \textit{Content}, \textit{Presentation}, \textit{Accessibility}, and \textit{Production}.
Among these issues, nearly half (\productratio) of them focus on \textit{Production};
\textit{Content}, \textit{Accessibility}, and \textit{Presentation} take \conratio, \accratio, and \presentratio, respectively.
We find that: 1) RN producers are more likely to miss information than to include incorrect information, especially for breaking changes; 2) improper layout may bury important information and confuse users; 3) many users find RNs inaccessible due to link deterioration, lack of notification, and obfuscate RN locations; 4) automating and regulating RN production remains challenging despite the great needs of RN producers.
Our taxonomy not only pictures a roadmap to improve RN production in practice but also reveals interesting future research directions for automating RN production.
\end{abstract}




\maketitle
\section{Introduction}
\label{sec:introdution}

When releasing a new software version, developers often produce a \textbf{release note} (RN) which summarizes main changes in the software since its previous release~\cite{moreno2017arena:}. 
RNs serve as means of communication between the software and its users~\cite{bi2020empirical}.
Consulting RNs is considered as an essential best practice when upgrading software~\cite{Converge23:online}.
Users typically use RNs to comprehend: 
1) potentially beneficial changes, such as bug fixes, enhancements, new features, to help them decide whether to upgrade to the new release;
2) potentially interrupting changes, along with guidance for migration or mitigation.
Besides, internal developers use RNs to formally document development progress and plans for the next release~\cite{bi2020empirical}.\footnote{
Note that the term ``release note'' often refers to the documentation that refine and summarize change logs. 
However, in practice many software projects directly use change logs as their release notes, so some results in our paper refer to both.

Throughout this paper, we use the term ``user'' to refer to anyone reading RNs or referring RNs for their tasks. 
A user can be an internal developer, a downstream developer, or a software end user.
} 

For large software projects, the production of RNs is both time-consuming and error-prone.
The survey by Moreno et al.~\cite{moreno2017arena:} found that \textit{``creating a release note by hand is a difficult and effort-prone activity that can take up to eight hours''}.
The tight deadlines in agile software development may even tempt developers to reduce effort put into RNs~\cite{Noreleas28:online}. 
Consequently, the produced RNs may be of low quality (bad organization, missing important changes, etc.), which brings various problems to software users.
However, previous researches~\cite{abebe2016empirical, moreno2017arena:, bi2020empirical, yang2021empirical} mainly focus on categorizing RN content and automated RN generation, while a systematic understanding of real RN issues in practice (\ie how RNs go wrong or fail to meet users' expectations) is still lacking.
Such an understanding can help formulate best practices and reveal important future research directions for automating and regulating RN production.
Therefore, to bridge the knowledge gap, we ask the following research question (\textbf{RQ}): \textit{What are the RN issues faced by developers?}

To answer this RQ, we collect \dataset RN-related GitHub issues from GHArchive~\cite{GHArchiv56:online}
and build a comprehensive taxonomy of these issues using multiple rounds of open coding.
The taxonomy is further validated through semi-structured interviews.
The final taxonomy consists of four main dimensions: \textbf{Content (\con)}, \textbf{Presentation (\present)}, \textbf{Accessibility (\acc)}, and \textbf{Production (\product)}, that reveals the challenges of using RNs and therefore the problems of producing RNs.
To the best of our knowledge, this is the \textit{first} paper that provides such a taxonomy.

Based on our taxonomy, we derive a practitioner-oriented checklist for RN production, which involves the selection of appropriate content, organization, and writing style for RNs.
We additionally provide recommendations for regulating RN production and ensuring RN completeness.
Finally, we identify open research challenges, which can benefit the automation of RN production and testing of RN completeness/correctness in practice.
We provide a replication package at \url{https://doi.org/10.6084/m9.figshare.18777650}.




\section{Background and Related Work}
\label{sec:related_work}


In the early years of software development, software products are often released ``once and for all'' with no modifications after the initial release.
However, successful software inevitably evolves into new versions.
When a new version needs to be released, documentation for explaining changes in this version, \ie Release Note (RN), emerges as a natural requirement.
Although we cannot precisely trace the history of the earliest RNs, the term ``release note'' has at least been used in the software industry since the 1980s~\cite{earliest_RN}.

From the beginning of the 21st century, the movement toward agile software development advocates ``release early, release often'' so that a tight feedback loop between developers and users can be created~\cite{olsson2014opinions}.
Consequently, the required effort to manage changes between consecutive software versions has significantly increased. 
Then, software projects begin to formulate systematic agendas for software release management, in which RNs are perhaps the most important kind of documentation~\cite{aghajani2020software}.
Nowadays, complex software systems such as Firefox have to deal with a tremendous amount (up to thousands) of patches during each release cycle, which creates a formidable challenge in tracking changes to be included in a RN and producing the final RN.
For Firefox, the Mozilla team defines a systematic process, including workflows, conventions, and automated tooling, to support the creation of RNs~\cite{mozilla_cycle:online}.


Meanwhile, RNs remain an understudied research topic.
Early studies only use RNs as a data source for understanding other software maintenance and evolution topics~\cite{yu2009mining,alali2008s,maalej2010can,shihab2013studying}.
It is not until the recent decade do researchers begin to study RNs themselves with two main fronts: empirical studies for understanding RN practices and approaches for automated RN generation.

\vspace{-2mm}
\subsection{Understanding Release Note Practices}
\label{sec:related_nature}


Moreno et al.~\cite{moreno2017arena:} manually analyze 1,000 RNs from 58 industrial and open source projects.
They identify 17 common change types in RNs, such as fixed bugs, new features, and new code components.
Similarly, Abebe et al.~\cite{abebe2016empirical} manually analyze 85 RNs from 15 software projects and identified six types of information: title, system overview, resource requirement, installation, addressed issues, and caveat. 
Bi et al.~\cite{bi2020empirical} study the characteristics of 32,425 RNs from 1,000 GitHub projects. 
They classify common RN content into eight topics including issues fixed, new features, system internal changes, etc. 
They find that RN content significantly differs across software in different domains, \eg for application software and system software, new features are most frequently documented. 
They further uncover discrepancies between RN producers and users through interviews and surveys.
However, it is still unclear \textit{what content tends to go wrong in RNs}, which may have a different distribution.

The nature of RN is also discussed in some work related to software documentation. 
Aghajani et al.~\cite{aghajani2020software} perform a survey with 146 developers to investigate what kind of documentation types are considered important in software development. 
They find that although the majority of developers consider RNs and change logs as important, their absence is also among their frequently encountered issues. 
Developers also suggest including documentation such as RNs as mandatory items in the release checklist.

Despite the discrepancies between RN producers and users as identified by Bi et al.~\cite{bi2020empirical}, we still lack a comprehensive empirical understanding of real issues in RN production and usage.
To the best of our knowledge, this is the \textit{first} paper toward this direction, and our taxonomy provides a significant amount of new empirical evidence for improving RN production in practice.

\subsection{Automating Release Note Production}
\label{sec:related_Generation}

Since producing RNs is both important and effort-prone, developers naturally begin to explore ways to automate this process.
For software projects managed via a version control system (VCS), 
the most straightforward way of producing a RN is to aggregate all changes from the VCS (\eg aggregating all commit messages from Git).
However, such simple way of automation comes with severe drawbacks, as noted by the OpenStack documentation:

\textit{``Release notes are not meant to be a replacement for git commit messages. They should focus on the impact for the user and make that understandable, even for people who do not know the full technical context for the patch or project''}~\cite{Openstack_management:online}.

To facilitate the production of high quality RNs while reducing manual effort, many open-source projects begin to adopt tools for automated RN generation, including Semantic Release~\cite{semantic-release:online} 
($\sim$14k stars),  github-changelog-generator~\cite{github_changelog_generator} ($\sim$6k stars), Release It~\cite{releaseit:online} ($\sim$4k stars), Release Drafter~\cite{releasedrafter:online} ($\sim$2k stars), etc.
All tools make the assumption that every software change should be documented using predefined templates or labels so that they can generate RNs based on predefined rules.
For example, Semantic Release requires developers to write commit messages in the format specified by Angular Commit Message Conventions~\cite{angularj59:online} with eight types of predefined changes. 
These tools are generally designed to be easily extensible and configurable to fit the needs of different projects.
Even if some automation is adopted, it is still common to post edit the RNs to summarize changes, highlight, or intrigue readers, etc.

To improve the state of practice, researchers have proposed novel approaches for automated RN generation.
Klepper et al.~\cite{klepper2016semi} propose a semi-automated RN generation tool which extracts change descriptions from issue trackers and organizes them by labels to meet the need of a specific audience. 
Moreno et al.~\cite{moreno2017arena:} propose a fully automated RN generation tool, ARENA, which integrates both changes from VCS and rationales for each change from issue trackers into RNs with predefined change categories.
Nath et al.~\cite{nath2021towards} propose to generate RNs from commit messages and pull requests using text summarization and word embedding techniques.
Jiang et al.~\cite{jiang2021deeprelease} propose a language-agnostic approach to produce RNs from pull request text using deep learning.

While several automated approaches have been proposed by researchers, we are still not aware of any wide industrial adoption, indicating potential discrepancies between research and practice.
Our work complements existing effort on RN automation by summarizing best automation practices and reveal future research directions for improving automated tools.

\vspace{-1mm}
\section{Methodology}
\label{sec:Methodology}

\subsection{Data Collection}

In this study, we choose to analyze GitHub issues, which developers use to track ideas, provide feedback, report bugs, and initiate discussions~\cite{Aboutiss_about-issues:online}. 
We favor GitHub issues over Stack Overflow questions because GitHub issues contain more information such as reports and discussions among developers and provide concrete examples about how RNs fail, apart from developers' opinions.

\vspace{-1mm}
\subsubsection{Mining GitHub}
GitHub is one of the most popular social coding platforms and provides access control and several collaboration features such as bug tracking, feature requests, task management for every project. It is a commonly used data source for exploring software issues in previous works~\cite{coelho2015unveiling, humbatova2020taxonomy}.
To this end, we use the GHArchive dataset~\cite{GHArchiv56:online} to collect all GitHub issues that:
1) have activities (at least one \Code{IssueEvent} in GHArchive) between January 2021 and June 2021;
2) contain the keyword ``release note'' in their titles. 
We only include the latest GitHub issues (with activities in 2021) because we observe that RN practices are rapidly changing in open-source communities, and thus data timeliness is vital.
For example, Bi et al.~\cite{bi2020empirical} report that developers do not use automated RN generation tools while the number of automated RN generation tools is gaining increasing popularity recently (Section~\ref{sec:related_Generation}).
This initial selection results in \dataset issues from 1,019 repositories.

\vspace{-1mm}
\subsubsection{Refining Dataset}
\label{ss:refining-dataset}

\begin{table}[!t]
\small
\setlength{\tabcolsep}{3pt}
\caption{Repository Statistics of the Final Issue Dataset}
\vspace{-3mm}
\label{tab:repos}
    \begin{threeparttable}
    \begin{tabular}{lrrrrrr}  
    \toprule   
          & Median  & Mean & Std. & Distribution\tnote{$*$}\\  
    \midrule  
        Age (in Days)      & 1,483.00 & 1.676.36 & 1,058.27 &
        \adjustimage{height=0.3cm,valign=m}{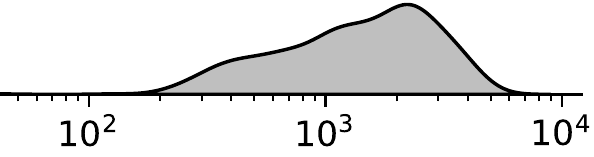} \\ 
        \# of Commits      & 1,053.00 & 7,357.25 & 36,020.93 & 
        \adjustimage{height=0.3cm,valign=m}{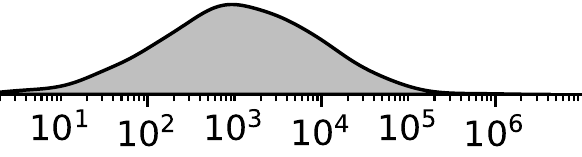} \\
        \# of Stars        & 188.00   & 4,321.06 & 13,536.23 &
        \adjustimage{height=0.3cm,valign=m}{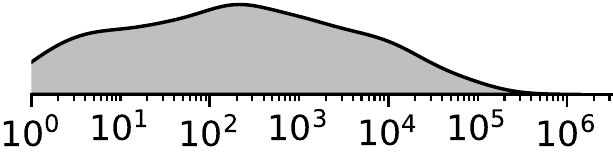} \\ 
        \# of Contributors & 28.00    & 88.23 & 122.39 &
        \adjustimage{height=0.3cm,valign=m}{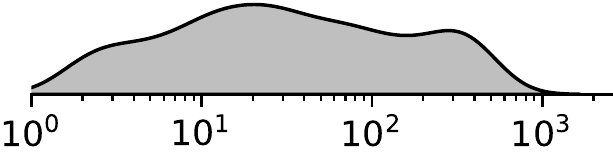} \\ 
        \# of Forks        & 69.50    & 1,024.87 & 3,430.94 &
        \adjustimage{height=0.3cm,valign=m}{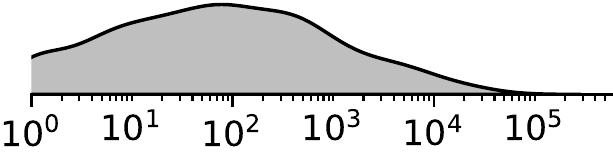} \\ 
        \# of Issues       & 53.00    & 426.57 & 2,517.74 &
        \adjustimage{height=0.3cm,valign=m}{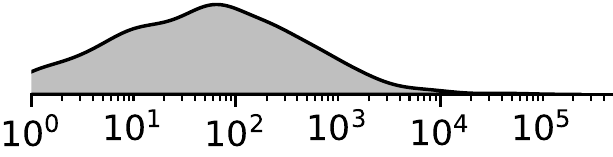} \\ 
        \# of PRs          & 5.00     & 36.22 & 188.31 &
        \adjustimage{height=0.3cm,valign=m}{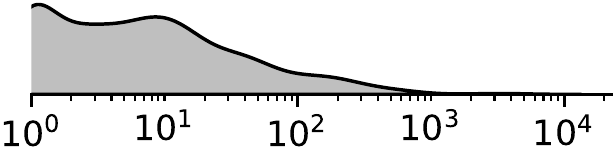} \\ 
        \# of Releases     & 13.00    & 60.50 & 250.59 &
        \adjustimage{height=0.3cm,valign=m}{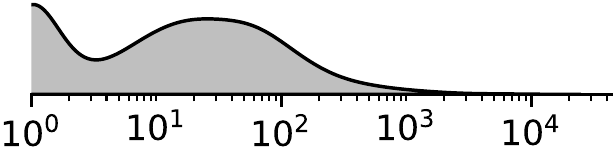} \\ 
        \bottomrule  
    \end{tabular}
    \begin{tablenotes}
    \footnotesize
    \item[$*$] We increment all values by one to plot the distribution in log-scale.
    \end{tablenotes}
    \end{threeparttable}
\vspace{-3mm}
\end{table}

Two authors (named as inspectors), both with over six years of software development experience, further read all the issues jointly to refine the final dataset.
The inspectors browse through the GitHub issue pages of the all collected issues together as an initial familiarization of the dataset and exclude \falsedata issues that are not related to certain problems in RNs (i.e., \textit{False Positives}), including the following cases:
\vspace{-0.5mm}
\begin{itemize}[label=\itembullet,leftmargin=10pt]
    \item \textit{Release Statements (\statement)}: The issue is only an official announcement of a release or a release note.
    \item \textit{Non-Informative (\noninfo)}: The issue contains too little information (e.g., only a few words in title and description) to be understood by the inspectors.
    \item \textit{Irrelevant (\irr)}: The issue happens to have the keyword ``release note'' in its title but actually refers to a problem not related to RNs.
    \item \textit{Unreachable (\unreach)}: The issue is no longer available on GitHub (\eg the repository is deleted or made private, the issue is deleted, etc.).
    \item \textit{Non-English (\noneng)}: The issue contains non-English text and is not understandable by the inspectors.
    \item \textit{Mistake (\mistake)}: The issue reporter misunderstands the RN and reports a non-existent problem.
\end{itemize}
\vspace{-0.5mm}

The final dataset for our study consists of \finaldataset issues from 722 repositories.
The repository statistics are summarized in Table~\ref{tab:repos}, 
where we can observe that most issues come from repositories with long development history, high popularity, and sufficient development activities.\footnote{The long tail distributions of most metrics are expected and common in mining software repository datasets~\cite{he2021large, zhang2019companies}.}
In fact, given that only software with a sufficiently large user base may consider writing RNs or have users reporting issues for RNs, it is natural that almost all of these issues come from mature software repositories.
The size of our dataset is comparable to and even larger than similar software engineering studies that conduct qualitative manual analysis on text (e.g., studies on Stack Overflow posts and patch descriptions \cite{aghajani2019software,beyer2018automatically,zhang2019empirical,chen2020comprehensive, tan2019communicate}).

\vspace{-1mm}
\subsection{Analysis Method}
\label{ss:analysis_method}

\begin{table}[!t]
\small
\centering
\caption{Statistics for Each Round of Manual Labeling}
\vspace{-3mm}
\label{tab:labelround}
    \begin{threeparttable}[t]
    \begin{tabular}{lrrrrrr}  
    \toprule
    Round    & 1  & 2 &  3& 4 & 5\tnote{$*$} &Total\\
    \midrule
    Analyzed              & 273     & 212   & 212     & 212    & 90  & \finaldataset\\
    Cohen’s Kappa & -     & 0.78 & 0.86   & 0.87  &  - & -\\
    \midrule
    Newly Added\\
    ~-~\#Dimensions     &3  & 1  &0  & 0   &  0 & 4\\
    ~-~\#Categories     &5  & 2  &0  & 0   &  0 & 7\\
    ~-~\#Subcategories  &7  & 4  &1  & 0   &  0 & 12\\
    ~-~\#Leaf Nodes     &48 & 7  &8  & 2   &  -3 & 62\\
    \bottomrule  
    \end{tabular}
    \begin{tablenotes}
    \footnotesize
    \item[$*$] The fifth round samples issues from previous four rounds (Section~\ref{sec:interview}, \ref{sec:reproducibility}).
    \end{tablenotes}
   \end{threeparttable}
\vspace{-5mm}
\end{table}

For the final \finaldataset RN-related issues, we follow an open coding procedure to inductively create the dimensions, categories, subcategories, and leaf nodes of our taxonomy in a bottom-up way~\cite{seaman1999qualitative}.
Similar to previous works~\cite{chen2020comprehensive, humbatova2020taxonomy}, our procedure of taxonomy construction consists of four steps: pilot construction, extended construction, developer interview, and reproducibility verification.
The four steps are integrated with a five-round labeling process and the statistics for each round of labeling are summarized in Table~\ref{tab:labelround}.

\vspace{-1mm}
\subsubsection{Pilot Construction.}
\label{sec:pilot-construction}
We randomly sample 30\% (273) of the \finaldataset issues for a pilot construction of the taxonomy in the first round with two stages. 
The inspectors mentioned in Section~\ref{ss:refining-dataset} independently analyze the underlying RN problems behind the sampled issues. 
In the first stage, the inspectors aim to be familiar with RNs' issues. 
They read and reread titles, descriptions, labels, and comments of each RN-related issue to understand its problems and intention.
Where necessary, they additionally check relevant code changes (i.e., pull requests/commits) and release notes that reveal the final solution adopted by project developers.
In the second stage, the inspectors assign short phrases as initial codes and record important information to indicate the problems and needs behind these issues.
If an issue is related to multiple problems and needs, e.g., the RN misses both new features and breaking changes, it will be assigned with multiple initial codes.
After the initial codes are generated, the inspectors proceed to group similar codes into categories, create a hierarchical taxonomy of RNs' issues, and assign issues to the taxonomy.
We include an additional arbitrator, who has several publications in top-tier software engineering venues and more than six years of software development experience, to mediate, discuss, and resolve any disagreement during taxonomy construction.
They continuously go back and forth between categories and issues to refine the taxonomy until the inspectors and the arbitrator finally approve all categories in the taxonomy. 

\vspace{-1mm}
\subsubsection{Extended Construction}
Based on the initial hierarchical taxonomy generated in Section~\ref{sec:pilot-construction}, the inspectors and the arbitrator iteratively conduct independent labeling, conflict resolution, and taxonomy refinement in the next three rounds.
In each round, two inspectors first independently label one-third of the remaining issues.
When they find issues that cannot be labeled in the current taxonomy, they add them to a temporary \textit{Pending} category. 
Then, the inspectors and the arbitrator organize a meeting to resolve labeling conflicts and determine whether new categories should be added for issues in the \textit{Pending} category.
After the taxonomy is refined, they update all previously labeled issues into the refined taxonomy and proceed to the next round.
Saturation is reached in the third round because we add only new leaf nodes (Table~\ref{tab:labelround}).
We finish labeling all the issues in the fourth round.
In the three rounds of extended construction, we use Cohen's Kappa ($\kappa$) to measure inter-rater agreement between two inspectors.
The $\kappa$ values are 0.78, 0.86, and 0.87, respectively, indicating increasing and high agreement between inspectors.

\vspace{-1mm}
\subsubsection{Developer Interview}
\label{sec:interview}

To validate our taxonomy with practitioners, we interview three industry software engineers from different large IT companies. 
They all have rich experience in publishing RNs with 1.5, 3, and 7 years of experience, respectively.

We opt for \textit{semi-structured} interviews.
Each of our interviews begins with the question: \textit{what issues have you encountered around RNs in your software development process?}
The purpose of this open-ended question is to see if our taxonomy covers the problems that developers usually encounter during development.
They each describe three, five, and two issues they encountered based on their own development experience.
Then, we present our taxonomy and direct them to specific categories of issues in our taxonomy, which enables them to recall other four previous issues.
All issues are covered by our taxonomy, indicating that our taxonomy has good coverage even within a different context (i.e., industry setting).

Then, we ask them to review and provide suggestions about our taxonomy.
They think our taxonomy is clear and informative, though some leaf nodes can be improved.
After discussion, we decide to merge seven leaf nodes into three leaf nodes and split one leaf node into two leaf nodes finally.
The interview time varied between 46 minutes and 2 hours. All interviews are conducted face to face with two authors (one is the leader and the other one asks additional questions when appropriate). 
The reason is that previous works~\cite{hove2005, humbatova2020taxonomy} show that participants talk much more when more than two interviewers conduct the interviews.

\vspace{-1mm}
\subsubsection{Reproducibility Verification}
\label{sec:reproducibility}

One problem remaining with our taxonomy is reproducibility because we intertwine taxonomy construction with independent labeling.
This is hard to avoid because the taxonomy is too complex to be precisely defined in one or two rounds.
Although we maintain a code book during the process, it is still unclear whether others can reproduce the taxonomy using the same code book.
Therefore, we invite two interviewees and one additional Ph.D. candidate to label issues using our code book.
Each of them is assigned 30 different issues and they return their results after 3 days.\footnote{We do not assign more because inspecting, comprehending, and labeling issues takes significant time and energy which they lack to label more.}
Compared with our own results, the $\kappa$ values are 0.93, 0.89, and 0.86, respectively, which also indicates a high agreement and thus good reproducibility.

Our final taxonomy includes four dimensions, seven categories, 12 subcategories, and 62 leaf nodes.
The entire manual construction process takes over two months to finish.

\vspace{-1mm}
\section{Results}
\label{sec:result}

\begin{figure*}[!ht]
  \centering
  \includegraphics[width=\linewidth]{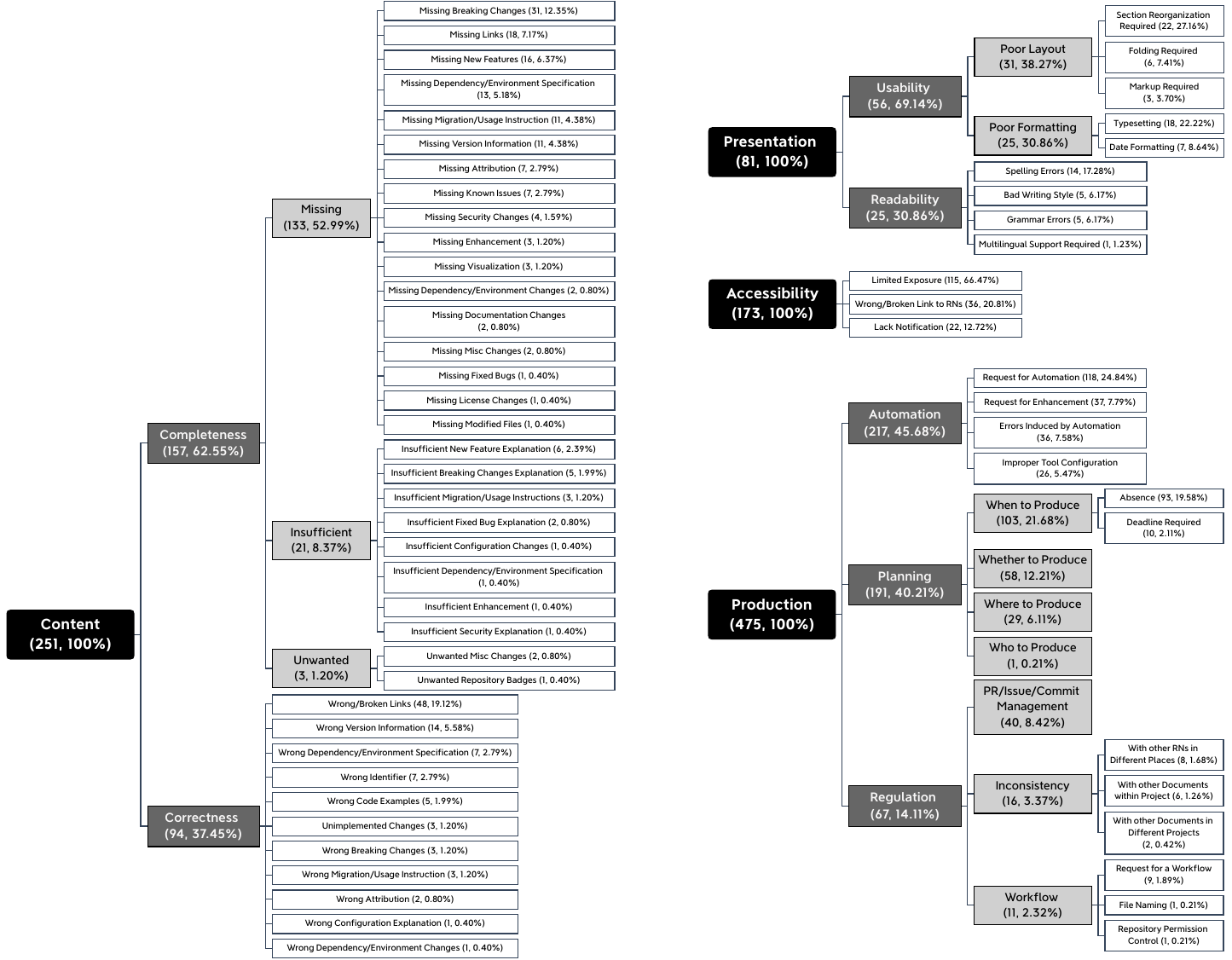}
  \vspace{-7mm}
  \caption{
  The Taxonomy of Release Note Issues. (\ColorRule[codeblack]{12pt}{6pt}) Represents Dimensions, (\ColorRule[codedeepgrey]{12pt}{6pt}) Represents Categories, (\ColorRule[codegrey]{12pt}{6pt}) Represents Subcategories, and (\ColorRule[backcolour]{12pt}{6pt}) Represents Leaf Nodes.
  }
  \label{fig:taxonomy}
  \vspace{-4mm}
\end{figure*}

Figure~\ref{fig:taxonomy} 
illustrates the hierarchical taxonomy of RN issues. 
We group all these issues into four \textbf{dimensions}:
\begin{enumerate}
    \item \textbf{Content}: What information should RNs convey?
    \item \textbf{Presentation}: How should RNs convey information?
    \item \textbf{Accessibility}: How to make RNs easily accessible?
    \item \textbf{Production}: In what way should RNs be produced?
\end{enumerate}
Each dimension is then hierarchically organized into \textbf{categories} (\eg \textit{\textsf{Completeness}}), \textbf{subcategories} (\eg \textit{Missing}), and \textbf{leaf nodes} (optional, \eg \textit{Missing Breaking Changes}). 
Figure~\ref{fig:taxonomy} also shows the number of issues and percentages (within dimension) for all dimensions, categories, subcategories, and leaf nodes in the taxonomy.
In the remainder of this section, we will describe our taxonomy with representative examples. 

\vspace{-1mm}
\subsection{Content}
\label{sec:content}

In total, \connum issues discuss the \textbf{Content} of RNs, \ie what information should RNs convey. 
Issues from the \textbf{Content} dimension can help better understand 1) what common mistakes developers often make when producing RNs, 2) what typical users would expect from RNs, and 3) what purposes RNs should serve as one kind of project documentation.
This dimension consists of two categories: \textit{\textsf{Completeness}} and  \textit{\textsf{Correctness}}.

\subsubsection{Completeness (157, 62.55\%)}
\label{sec:completeness}
This category of issues concerns whether RNs contain both sufficient and necessary information required by users during software upgrades or required by internal developers for maintenance purposes.
It has three subcategories: \textit{Missing}, \textit{Insufficient}, and \textit{Unwanted}.


\myfancylabel\;\textit{Missing (133, 52.99\%)} subcategory refers to issues stating that some information perceived important by end users or internal developers is not included in the RN at all.
The most frequently missed information in RNs includes:
\begin{itemize}[label=\itembullet,leftmargin=10pt]
    \item \textit{Breaking Changes (31, 12.35\%)}:
        Such issues are predominant because end users directly encounter upgrade failures if they are not notified of breaking changes from reading RNs. 
        However, it can be difficult for RN producers to correctly locate and highlight breaking changes in RNs.
        For example, a developer from \Code{mongoose} notes that \textit{(the new version) has many errors, and fixing them is not just changing a function/field name, because function parameters/semantics have also changed} because of the \textit{undocumented breaking changes from v6 to v7}~\cite{mongoose_1271_issues:online}.
    \item \textit{Links (18, 7.17\%):} 
        In these issues, developers ask for links to external materials (\eg related PR/issues/commits, usage guides, CVEs, etc.) to better understand information conveyed in RNs.
    \item \textit{New Features (16, 6.37\%):} 
        Some implemented new features may be ignored in RNs, and (other) developers open issues in need of documenting their contributions.
    \item \textit{Dependency/Environment Specification (13, 5.18\%):}
        Undocumented dependency or environment specification may also accidentally break clients when users upgrade to new versions.
    \item \textit{Migration/Usage Instruction (11, 4.38\%):}
        Some developers ask for migration or usage instructions in RNs to help them understand the impact of breaking changes and upgrade their client code.
    \item \textit{Version Information (11, 4.38\%):} 
        Some developers open issues to discuss adding version information in RNs, \eg release date, version number \& name, checksum, and release status (draft or final) for easy reference to specific releases.
    \item \textit{Attribution (7, 2.79\%):}
        Some issues are opened by repository members to discuss missing attribution to certain participants (\eg contributors, funders, commenters, reviewers, etc.).
        As stated by a maintainer of \Code{coq}, \textit{in open source software, it is very important to give credit}~\cite{Howtopre_7058_issue:online}. 
    \item \textit{Known Issues (7, 2.79\%):} 
        Several issues mention that specific unsolved issues should be included in RNs to alert end users, \eg including a \Code{NullPointerException} crash and its workaround in the corresponding RN of \Code{NuGet}~\cite{Addknown_2410_issue:online}. 
\end{itemize}
Other kinds of information may also be reported as missing, though less frequently, including notification of security changes, enhancements, visualization (additional diagrams or plots), documentation changes, fixed bugs, license changes, modified files, etc.

\myfancylabel\;Issues in the \textit{Insufficient (21, 8.37\%)} subcategory arise because certain information related to important changes is not sufficiently detailed for users to understand.
Two kinds of explanations are most likely to be insufficient in RNs:
\begin{itemize}[label=\itembullet, leftmargin=10pt]
    \item \textit{New Feature Explanation (6, 2.39\%):} Developers tend to ask for more information about unfamiliar new features if they intend to use them after upgrading. 
    For example, \Code{Keras} 2.0 renames \Code{samples\_per\_epoch} to \Code{steps\_per\_epoch} in \Code{fit\_generator()} but its RN fails to mention additional changes in parameter semantics, which confuses downstream developers~\cite{keras_11517_issue:online}.
    \item \textit{Breaking Change Explanation (5, 1.99\%)}: Developers also ask for more clarification about changes that may break downstream code. We observe a vivid example in \Code{numpy} where a developer opens an issue to argue that \textit{we should try to improve the release notes (and probably warnings) for the \Code{np.int} and other python alias deprecations}~\cite{DOCConsi_17977_issue:online}.
\end{itemize}
Other insufficiently explained information include: migration/usage instructions, fixed bugs, configuration changes, dependency/environment specification, enhancements, and security. 

\myfancylabel\;Interestingly, three issues care about \textit{Unwanted} information in RNs, but they are likely to be only occasional.
Two issues state that only critical/developer-impacting changes should go in release notes instead of listing all miscellaneous changes~\cite{Updateau_1873_issue:online}, while the other issue~\cite{moodle-tool_ribbons_3_issue:online} mentions that repository badges should not occur in release notes.


\subsubsection{Correctness (94, 37.45\%)}
\label{sec:correctness}
This category means that information described in RNs conveys inaccurate information.

\myfancylabel\;Contrary to our intuition, the majority is \textit{Wrong/Broken Links (48, 19.12\%)}, which refers to cases where links in RNs cannot be opened or direct to an incorrect page.
Most links should point to other kinds of documentation, \eg user guide, for the elaboration of changes in RNs; others are expected to point to related PR/issue/commit, project main branch, the homepage of other projects, RNs of sibling projects, files for download, etc.
These links are supposed to supplement information, but they tend to deteriorate over time, 
which causes poor reading experience for RN readers.

\myfancylabel\;Moreover, 14 issues are related to \textit{Wrong Version Information (14, 5.58\%)}, including version number/name, version date, checksum, and most of which are caused by copy-pasting from previous RNs~\cite{nushell_3238_issue:online,babel_12961_issue:online,WebView2Feedback_882_issue:online}.
Other kinds of change descriptions that can go wrong include identifier, dependency/environment specification, code examples, breaking changes, migration/usage instructions, unimplemented changes,
attribution, dependency/environment changes, explanation of configuration, etc.


\vspace{-2mm}
\begin{result-rq}{Summary for \textbf{Content}:}
Nearly two-thirds (62.55\%) of issues within this dimension concerns \textit{\textsf{Completeness}}, while only about one-third (37.45\%) concerns \textit{\textsf{Correctness}}. 
Developers are most likely to 1) report wrong/broken links in RNs (19.12\%), which annoyingly prevent them from accessing supplementary information, 
and 2) missing breaking changes (12.35\%), which may mislead users and incur severe consequences after upgrading (\eg crash).
\end{result-rq}
\vspace{-2mm}

\subsection{Presentation}
\label{sec:results-presentation}

\presentnum issues are related to \textbf{Presentation}, with two categories: \textit{\textsf{Readability}} and \textit{\textsf{Usability}}. 
Issues from the \textbf{Presentation} dimension help reveal how information should be organized, formatted, highlighted, visualized, and phrased in an RN, so that different RN users can make use of the RN for their purposes with maximum efficiency.

\subsubsection{Usability (56, 69.14\%)} 
\label{sec:usablity}
This category refers to the degree to which users can use RNs to achieve their objectives effectively.

\myfancylabel\;More than half of the issues (31, 38.27\%) in this category are related to
\textit{Poor Layout}, which means the changes are not clearly organized in RNs.
Since different stakeholders may be interested in different kinds of information, RNs need to have a proper layout for them to quickly locate the information they want~\cite{bi2020empirical,abebe2016empirical}.
On the other hand, poorly organized RNs may increase the time needed for users to grab valuable information, annoy readers~\cite{platform_2178_issue:online},
bury good features~\cite{eos_9903_issue:online},
and cause important changes to be missed by impacted users.
For example, \Code{Electron} lists two API deprecations under the ``other'' section in the v12.0.0 RN by mistake, which makes the deprecations easily overlooked~\cite{electron_28375_issue:online}. 
Developers make the following suggestions in these issues for improving RN layout:
\vspace{-0.5mm}
\begin{itemize}[label=\itembullet,leftmargin=10pt]

    \item \textit{Section Reorganization (22, 27.16\%)}: Reorganize changes into a separate section if they are concerned by a specific audience, \eg a separate section for database operators in the RNs of \Code{CockroachDB}~\cite{cockroach_57898_issue:online}.
    \item \textit{Folding (6, 7.41\%)}: Shorten RNs and fold a lengthy list of details, using detail/summary tags provided by GitHub Release Page~\cite{prisma_5913_issue:online}, 
    HTML, or Markdown features~\cite{platform_2178_issue:online}.
    \item \textit{Markup (3, 3.70\%)}: Use markups (\eg icons or emojis) to highlight breaking changes.
\end{itemize}
\vspace{-0.5mm}



\myfancylabel\;Other issues arise from \textit{Poor Formatting (25, 30.86\%)}:
\vspace{-0.5mm}
\begin{itemize}[label=\itembullet,leftmargin=10pt]
    \item \textit{Typesetting (18, 22.22\%)}: Most of these issues are caused by misusing syntax of markup languages (\eg HTML) and usually lead to abnormal display, \eg failing to display list due to missing HTML  linebreaks~\cite{Convertr38:online}.
    \item \textit{Date Formatting (7, 8.64\%)}: Some date formats can cause ambiguities to people in different geographical regions~\cite{cornerstone_1990_issue:online}.
    
    

\end{itemize}
\vspace{-0.5mm}
\subsubsection{Readability (25, 30.86\%)}
\label{sec:readability}
This category of issues concerns whether the RN is easy to read, including three subcategories: \textit{Spelling Errors (14, 17.28\%)}, \textit{Grammar Errors (5, 6.17\%)}, \textit{Bad Writing Style (5, 6.17\%)}, and \textit{Multilingual Support Required (1, 1.23\%)}.
Although fixing grammar errors and spelling errors are easy, they may be hard to notice, especially for technical terms (\eg MACs and Macs~\cite{OfficeDocs-OfficeUpdates_300_issue:online}).
Also, certain writing style can make RNs clearer and more easily understandable, such as describing what happens after a bug is fixed instead of what used to happen~\cite{cloud-docs_18_issue:online}.
One issue asks for multilingual support which helps more users understand RNs and enables product adaptation to a broader market~\cite{AddMulti50:online}.




\vspace{-2mm}
\begin{result-rq}{Summary for \textbf{Presentation}:}
The majority of issues (69.14\%) within this dimension concerns \textit{\textsf{Usability}}, especially poor layout, which may bury important information and lead to end users' misjudgement. Developers propose various solutions to alleviate this problem, including section reorganization (27.16\%), folding (7.41\%), and use of markups (3.70\%). Other presentation issues concern \textit{\textsf{Readability}} (30.86\%), such as spelling, writing style, etc.
\end{result-rq}
\vspace{-2mm}

\subsection{Accessibility}
\label{sec:accessibility}

\accnum issues are related to \textbf{Accessibility}, \ie how to make RNs accessible to a broad audience, with three categories: \textit{\textsf{Limited Exposure}}, \textit{\textsf{Wrong/Broken Link to RNs}}, and \textit{\textsf{Lack Notification}}.
Issues in this dimension thus reveal how a software project should distribute their RNs, maintain links, and notify their users.

\subsubsection{Limited Exposure (115, 66.47\%)} 
\label{sec:exposure}
Issues under this subcategory express either difficulty in finding RNs or expectation of more available ways to access RNs.
The former case happens when RNs are placed in obscure locations, \eg files with an unconventional name or in a deeply nested directory~\cite{hedgedoc.github.io_59_issue:online}. 
In the latter case, users suggest various locations to show RNs, \eg \textit{Can we get a page which explains the features and the release numbers in each of the releases of Teams Clients? If you have one, we can not find it}~\cite{SqlClient_1123_issue:online}.

\subsubsection{Wrong/Broken Link to RNs (36, 20.81\%)}
\label{sec:wrongtoRNs}
\textit{Missing Links} and \textit{Wrong/Broken Links} under the \textbf{Content} dimension describe cases where links in RNs are broken or wrong (see Section~\ref{sec:completeness} and Section~\ref{sec:correctness}). 
In addition, many issues report that external links that suppose to point to RNs themselves (in project website, etc.) may be broken. 
For example:
\begin{enumerate}[leftmargin=15pt] 
    \item \textit{The section of the front page of this repo ``Links to release notes'' is full of dead links}~\cite{HOK-Revit-Addins_194_issue:online}.
    \item \textit{Links to the Agent release notes from the APM docs left nav are returning 404s}~\cite{docs-website_758_issue:online}.
\end{enumerate}


\subsubsection{Lack Notification (22, 12.72\%)}
\label{sec:lack-notification}

The concerns expressed by these issues are twofold: 
1) whether a certain medium should be adopted to notify users and publicize RNs, and 
2) whether current ways of notification should be improved.
For example, several issues mention the use of RSS feeds to notify new releases.
In another case, a user complains:
\textit{Currently, the release notes for an updated version only show after the new version is installed. Basically, it is preferable to know in advance what changes are made to the app before its downloaded and installed}~\cite{sublime_merge_1097_issue:online}.


For improving RN accessibility, developers in our studied issues suggest the following locations for putting RNs:
\vspace{-0.5mm}
\begin{itemize}[label=\itembullet,leftmargin=10pt]
    \item \textit{GitHub Release Pages}: GitHub provides a dedicated page to display the release history. 
    Many issues show that developers often check GitHub Release Pages first when searching for RNs because they consider GitHub Release Pages as the most intuitive location for releases and RNs.
    As stated by a developer: \textit{
    From an engineering point of view, having release notes published on GitHub is ideal, this is our source of truth}~\cite{docs-website_2127_issue:online}. 
    \item \textit{Project Websites}: Developers also expect project websites as RN management centers providing links to RNs for each release.
    Developers also suggest using a specific URL for the latest RN~\cite{3drepo.io_2502_issue:online}.
    \item \textit{Files in Repositories}: Some developers think that a RN file in the repository (the root directory or the \Code{doc/} folder) is more important than storing RNs on GitHub Release Page~\cite{dcrwallet_1966_issue:online}.
    A RN file in repository makes the repository more self-contained (not depending on GitHub) and allows the usage of collaborative editing tools like Git~\cite{woodwork_808_issue:online}. 
    The OpenStack community also requires that its projects must include RN files to record version changes and believe that this way can work on multiple patches simultaneously and reduce merge conflicts~\cite{reno_latest:online}.
    \item \textit{Apps}: Application software can provide buttons and links to access the latest RN, \eg an `about' button~\cite{woodwork_808_issue:online}.
    RN notifications can also be displayed when a new version is released~\cite{chef-workstation-app_302_issue:online}. 
    \item \textit{Instant Messaging Channels}: Such channels can be used to immediately deliver new RNs to subscribed end users, 
    \eg Slack~\cite{code-server_3193_issue:online}
    and Telegram~\cite{wayback_48_issue:online}.

\end{itemize}

\vspace{-2.5mm}
\begin{result-rq}{Summary for \textbf{Accessibility}:}
Users encounter a diverse range of difficulties in accessing RNs, including \textit{\textsf{Limited Exposure}} (66.47\%), \textit{\textsf{Wrong/Broken Links to RNs}} (20.81\%), and \textit{\textsf{Lack of Notification}} (12.72\%). 

\end{result-rq}
\vspace{-2mm}

\subsection{Production}
\label{sec:results-production}

\productnum issues fall into the \textbf{Production} dimension, \ie in what way RNs should be produced. 
Problems behind these issues can shed light on prospective automation approaches, improvement of existing tools, and design of better release processes.
This dimension consists of three categories: \textit{\textsf{Automation}}, \textit{\textsf{Planning}}, and \textit{\textsf{Regulation}}.

\vspace{-1mm}
\subsubsection{Automation (217, 45.68\%)} 
\label{sec:results-automation}

This category reflects four kinds of issues that developers frequently encounter on automated RN generation: \textit{Request for Automation (118, 24.84\%)}, \textit{Request for Enhancement (37, 7.79\%)}, \textit{Error Induced by Automation (36, 7.58\%)}, and \textit{Improper Tool Configuration (26, 5.47\%)}.


\myfancylabel\;\textit{Request for Automation (118, 24.84\%)}: 
More than half of issues in the \textit{\textsf{Automation}} category are opened for discussing whether some sort of automation should be used and what specific tools to adopt for managing and generating RNs.
As stated by a developer from \Code{spid-compliant-certificates-python}: \textit{Despite important, writing release notes is a very boring task...It would be nice to have them automatically generated every time a PR is merged}~\cite{spid-compliant-certificates-python_6_issue:online}.
Developers propose two main types of solutions for automation: 1) writing project-specific scripts to fill predefined templates and publish releases (e.g., ~\cite{rdf4j_2784_issue:online}),
and 2) adopting existing automated tools such as Semantic Release~\cite{semantic-release:online}, github-changelog-generator~\cite{github_changelog_generator}, Release It~\cite{releaseit:online}, and Release Drafter~\cite{releasedrafter:online}.
Although RN automation is a huge help in reducing manual toil, some developers express their concerns for full automation in their projects. They think that an automated workflow: 
1) requires prefixes or labels for classifying commits or PRs, which may burden the code review and CI complexity;
2) is not suitable for important versions, e.g., stable releases, which need manual editing for better readability. 

\myfancylabel\;\textit{Request for Enhancement (37, 7.79\%)}:
These issues reveal suggested improvement of automated generation tools and scripts by users.
Specifically, users mostly request for three kinds of support: 
1) automated generation of RNs for different branches~\cite{Addrelea66:online}; 
2) automated retrieval of related information from multiple repositories~\cite{Generate11:online}; 
3) automated supplement of details, e.g., CVEs~\cite{Addingse26:online}, attribution~\cite{Assignau28:online},
and PR comments~\cite{Addrelea44:online}.
There are also some specific needs from various scenarios.
For example, a member of \Code{CockroachDB} proposes an extension to the RN extraction script to support the amendment of past RNs with new commits~\cite{cockroach_42163_issue:online}. 
Another issue opened by a contributor of \Code{chef/automate}, reveals the limited support of combining multiple RNs when upgrading across multiple versions and asks for further improvement~\cite{automate_2141_issue:online}. 
Some developers suggest current tools to also add support to automate RN publishing.

\myfancylabel\;\textit{Errors Induced by Automation (36, 7.58\%)}:
These issues report defects of automated tools and scripts, which are diverse and largely tool/project-specific. 
There are two types of issues:
one is that these defects lead to unexpected RNs, e.g, wrong/missing content~\cite{genchang6:online}, repetition~\cite{hugo_83_pull:online}, 
and incorrect positions~\cite{WrongRel7:online}; 
the other one is that these defects affect the generation process failure, e.g., not generating RNs that exceed certain length~\cite{azure-pipelines-tasks_11922_issue:online,mapeo-mobile_581_issue:online}.
Among these issues, most of them are caused by defects of current tools rather than project-specific scripts.
Besides, these tools all have to fetch changes history from Git and several issues are caused by its complex mechanisms, such as branch control~\cite{Releasen28:online}, rebase~\cite{istio_31816_issue:online} and release tag~\cite{Releasen10:online}, which developers need to pay more attention to in design.


%



%

\myfancylabel\;\textit{Improper Tool Configuration (26, 5.47\%)}:
These issues usually arise from unfamiliarity with the tools, such as generating RNs without a template or by a wrong template.
Most of these issues are caused by misconfigured change scopes, e.g.,
the expected branch, 
version ranges~\cite{CICreate1:online},
certain types of changes~\cite{ReleaseN89:online}, and triggered conditions~\cite{Onlygene8:online}.
Besides, parameter misconfiguration is another common cause, including the construction of paths ~\cite{Fixrelea82:online}, repository names~\cite{Fixediss63:online}, environment variables~\cite{Unableto46:online}, etc.



\vspace{-1mm}
\subsubsection{Planning (191, 40.21\%)}
This category of issues 
has four subcategories: \textit{When to Produce (103, 21.68\%)}, \textit{Whether to Produce (58, 12.21\%)}, \textit{Where to Produce (29, 6.11\%)}, and \textit{Who to Produce (1, 0.21\%)}.



\myfancylabel\;\textit{When to Produce (103, 21.68\%)}:
Developers discuss two kinds of issues in this subcategory, \textit{Absence} and \textit{Deadline Required}. 
\vspace{-0.5mm}
\begin{itemize}[label=\itembullet,leftmargin=10pt]
\item In the former case of \textit{Absence (93, 19.58\%)}, projects do not provide RNs for all releases (i.e., some releases are missing RNs), which causes their users to open inquiry issues.
For example, the absence of RN for \Code{Recoil} confuses a user who says: \textit{I saw that version 0.1.3 has been published on npm, but I cannot find release notes anywhere, would be good to know about potential breaking changes, deprecations and new additions}~\cite{Arethere60:online}.
Besides, although most projects provide RNs for every version, some of them are released too late to be helpful which disappoints users~\cite{terraform-ls_533_issue:online}.

\item In the latter case of \textit{Deadline Required (10, 2.11\%)},
developers discuss how to produce RNs timely, \eg update RNs before a new version is released~\cite{ApplicationInsights-Java_1682_issue:online},
give a deadline for the RN~\cite{openoppstasks_376_issue:online},
and announce the adoption of a formal release cycle~\cite{cookiecutter-girder-4_45_issue:online}.
\end{itemize}

\myfancylabel\;\textit{Whether to Produce (58, 12.21\%)}:
This subcategory discusses the necessity of providing RNs. 
Some projects never provide RNs for informing changes in the new release.
Consequently, in some cases, users open issues because the lack of RNs directly leads to upgrade failures and frustration~\cite{sentry-laravel_260_issue:online}. 
They have to resort to various effort-prone methods to figure out changes from commit history, \eg using \Code{git diff} to show all code changes between two versions~\cite{kcache_79_issue:online}.
Although \Code{git log} can list all commit messages and ease the pain of figuring out changes to some extent, as stated by a member of \Code{Common-Workflow-Language}, \textit{This requires everyone to write the best possible git commit message and have very clean git histories. While people are capable of this, it is more work for contributors}~\cite{cwlviewer_328_issue:online}.
In other cases, internal developers open such issues as they notice RNs \textit{would help developers to precisely see what notable changes have been made between each release of the project}~\cite{cwlviewer_328_issue:online}.
However, not everyone agrees with providing a RN with each release, because they think the changes are only internal or too minor to be worth mentioning~\cite{itsabito3:online,addchang:online}. 
Other project maintainers acknowledge the necessity of RNs but they lack time for them~\cite{Releasen14:online}.

\myfancylabel\;Different from \textbf{Accessibility} issues, issues in the \textit{Where to Produce (29, 6.11\%)} subcategory concern where to collaboratively edit and store RN files. 
Although GitHub provides convenient release functionalities~\cite{about-releases:online} to help developers manage RNs, it currently does not support collaborative RN editing.
By contrast, many projects with a large team wish to distribute RN workload among team members so that RNs can be scalably produced.
As a result, most projects opt for adding RNs as files in the git repository so that multiple developers can be involved in RN production (e.g.,~\cite{dcrwallet_1966_issue:online}). 

\myfancylabel\; One special case mentions the lack of accountability in RN production and suggests someone should be responsible for it~\cite{MakeourR40:online}.

\vspace{-0.5mm}
\subsubsection{Regulation (67, 14.11\%)}
\label{ss:regulation}
\label{sec:regulation}
This category of issues refers to what regulations should be followed to simplify and ease the production of RNs. 
It covers three subcategories: \textit{PR/Issue/Commit Management (40, 8.42\%)}, \textit{Inconsistency (16, 3.37\%)}, and \textit{Workflow (11, 2.32\%)}. 

\myfancylabel\;\textit{PR/Issue/Commit Management (40, 8.42\%)}: This subcategory refers to issues discussing how to efficiently prepare (relevant) PRs, issues, and commits for RNs.
This procedure is usually time-consuming, especially for large projects.
For example, a member from \Code{pytorch/vision} complains that \textit{I wrote the release notes last week and we spent the vast majority of the time labeling the PRs} and suggests \textit{it'd be good to have a process that would make this faster}~\cite{vision_3351_issue:online}.
Some solutions emerge from the discussions in these issues. 
For commits, developers prefer to adopt a convention for writing \textit{structured commit messages} (e.g., Conventional Commits~\cite{ConventionalCommits:online}), so that changes (e.g, features, fixes, and breaking changes) in a commit can be documented in a machine-parsable way. 
For PRs, several large projects recommend labeling each PR with pre-defined labels.
In the case of \Code{pytorch/vision}, developers reach a consensus on categorizing each PR with labels describing affected components and changed types (\eg breaking changes and improvements)~\cite{vision_3351_issue:online}. 
For issues, many developers mention the use of GitHub milestones~\cite{about-milestones:online} for progress tracking.
Some projects create each milestone using version numbers and group issues into milestones~\cite{multi-tenancy_916_issue:online},
which reduces the scope of review when developers write RNs.

\myfancylabel\;\textit{Inconsistency (16, 3.37\%)}:
This subcategory refers to issues about inconsistencies between 1) RNs published in different places, 2) RNs and other documentation within project, and 3) RNs and documentation in other projects.
As revealed in the \textit{\textsf{Accessibility}} dimension, RNs are usually published in different places including, GitHub Release Page, project homepage, etc. 
However, developers sometimes neglect to maintain their consistency. 
For example, a user suggests that \textit{It'd be great to have a way to sync release notes in \url{docs.newrelic.com} by fetching the information from GitHub}~\cite{docs-website_2127_issue:online}. 
RNs can also easily become inconsistent with other documentation within project, \eg usage guides and READMEs. 
As an example of inconsistency between RNs and usage guides, a user complains that \textit{our documentation is horribly outdated} and calls for internal developers to \textit{go through all release notes and move all information that is not outdated and is missing from the documentation to the usage guide}~\cite{otki_2527_issue:online}.
Finally, RNs sometimes need to include changes or attribution information from closely related projects, which requires collaboration of developers from the related projects.

\myfancylabel\;\textit{Workflow (11, 2.32\%)}:
This subcategory refers to issues discussing formulation of RN production workflow or improvement on existing workflow.
Among the issues, nine are opened as a \textit{Request for a Workflow}. 
For example, a developer from \Code{mantid/mantidimaging} formulates a workflow as follows: \textit{Release notes should be continuously updated during development. 
Almost all pull requests should have an update to the relevant file and section in \Code{docs/release\_notes}. If the next release name is not yet chosen, this will be \Code{next.rts}, and renamed closer to release. When fixes are backported to a release branch, they can be added to the notes for that release, in an updates section}~\cite{mantidimaging_798_pull:online}.
One issue discusses \textit{File Naming} and suggests avoiding confusing RN naming format~\cite{Currentr40:online}.
Another issue discusses \textit{Repository Permission Control} where developers in \Code{kubernetes/test-infra} request to have the permission to collaboratively edit RNs. 
This project only allows very few members to have write access to RNs, causing others to \textit{ping someone with write access or the author of the parent PR to add the release note to the PR body}~\cite{test-infra_9098_issue:online}.

\begin{result-rq}{Summary for \textbf{Production}:}
Developers show a strong interest in \textit{\textsf{Automation}} (45.68\% of issues within this dimension), but automated tools/scripts may lack desired features, tend to induce errors, and are hard or error-prone to configure. Additionally, without proper \textit{\textsf{Planning}} (40.21\%), \eg release schedules and deadlines, users may be confused about the absence of RNs. 
Finally, \textit{\textsf{Regulation}} (14.11\%) of RN production, especially conventions for pull requests (PRs), issues, and commits (8.42\%), is vital for enabling efficient RN production in large software projects.
\end{result-rq}
\vspace{-2mm}

\section{Discussion}

\subsection{Implications}
\label{sec:implication}

\subsubsection{Comparison with Previous Work}

\begin{table*}[!t]
\small
\setlength{\tabcolsep}{3pt}
\caption{Comparison of Most Frequent RN Content in Different Taxonomies.}
\vspace{-3mm}
\label{tab:repos}
\begin{threeparttable}
    \begin{tabular}{llll}  
    \toprule   
        Moreno et al.~\cite{moreno2017arena:} & Bi et al.~\cite{bi2020empirical} & Ours (\textsf{Completeness})\tnote{$*$} & Ours (\textsf{Correctness})\tnote{$*$} \\
    \midrule  
        Fixed Bugs (90\%) & Issues Fixed (79.3\%) & Breaking Changes (22.93\%) & Links (51.06\%)\\
        New Features (46\%) & New Features (55.1\%) & New Features (14.01\%) & Version Information (14.89\%) \\
        New Code Components (43\%) & System Internal Changes (25.1\%) & Links (11.46\%)& Dependency Specifications (7.45\%) \\
        Modified Features (26\%) & Non-functional Requirements (10.3\%) & Dependency Specifications (8.92\%) & Identifiers (7.45\%) \\
        Refactoring Operations (21\%) & Documentation Updates (9.5\%) & Migration/Usage Instructions (8.92\%) & Code Examples (5.32\%) \\
    \bottomrule  
    \end{tabular}
    \begin{tablenotes}
    \footnotesize
    \item[$*$] The percentages here are different from Figure~\ref{fig:taxonomy} because the denominators are the total number of issues in the \textsf{Completeness} (157 issues) and the \textsf{Correctness} (94 issues) category, respectively. In the \textsf{Completeness} column, issues from \textit{Missing} and \textit{Insufficient} are merged.
    \end{tablenotes}
    \end{threeparttable}
\label{tab:comparison}
\vspace{-4mm}
\end{table*}

Since previous works categorize RN content into different taxonomies~\cite{moreno2017arena:, bi2020empirical}, it is not easy to perform a detailed comparison of our results with theirs (mapping results from different work can be a possibility for future studies).
We can still observe some interesting differences from the most frequently occurring RN content in different taxonomies (Table~\ref{tab:comparison}):
1) breaking changes and links are more likely to have issues but they are not listed as a major category in previous taxonomies;
2) new features and bug fixes are not likely to have issues even if they occur most frequently in previous taxonomies;
3) some information frequently desired by users are not mentioned in previous work, such as migration/usage instructions, code examples, and dependency specifications.
Our lens of observation sheds light on the most fragile parts of RNs untouched in previous taxonomies.

The taxonomy in our paper also extends the work of Bi et al.~\cite{bi2020empirical} with a significant amount of new empirical evidence and actionable implications. 
For example, they find in RQ2.2 that clear structure and the writing styles of release note documentation are vital.
We go one step further and identify concrete evidence on how structure and style impact users, which we further derive into actionable advice on how to write and organize RNs.

\vspace{-2mm}
\subsubsection{A Checklist for RN Production}
\label{ss:checklist}

Based on the results summarized in Section~\ref{sec:result}, we provide a checklist as follows.

\faCheckSquareO\;\textit{What Should be Included in RNs?}
We find that issues related to RN \textbf{Content} (Section~\ref{sec:content}) have different distribution compared with the most frequent RN content identified in previous works~\cite{abebe2016empirical, moreno2017arena:, bi2020empirical}, which indicates that some types of information are more likely to be missed or incorrect than others.
Therefore, we recommend RN producers to check whether the following eight kinds of \textit{changes} have been described in RNs: 
1) Breaking Changes, 
2) New Features,
3) Enhancements, 
4) Fixed Bugs, 
5) Documentation Changes, 
6) Dependency/Environment Changes, 
7) Security Changes,
and 8) License Changes. 
We also find that additional information that benefits better understanding and tracking of these changes, \eg links to corresponding PRs/issues/commits, is preferred by users.
We therefore recommend including, where necessary, the following eight kinds of \textit{explanatory information} in RNs: 
1) Links to Change-Related PRs, Issues, and Commits,
2) Guides (\eg upgrade, migration, or setup guides),
3) Code Examples,
4) Dependency/Environment Specification, 
5) Attributions (\eg authors, reviewers, commenters, etc.),
6) Explanation for Jargon-Heavy Descriptions,
7) Versioning Information (\eg release time, version name/number, setup package, etc), 
and 8) Known Issues.

\faCheckSquareO\;\textit{How to Ensure RNs' Completeness?} 
Our results from Section~\ref{sec:content} show that RNs are more frequently affected by \textit{\textsf{Completeness}} issues than \textit{\textsf{Correctness}} ones, \eg missing breaking changes, which indicates the importance of ensuring completeness in RN production.
Thus, applying completeness checks on RNs, \ie making sure all critical changes are listed, is strongly recommended.
However, our investigations reveal two main reasons for completeness issues:
1) lacking manpower or time to conduct thorough inspections; 
2) difficult for a limited number of developers to understand all changes between versions.
While automated tools for checking RN completeness are still lacking, we locate several practices suggested by issue participants that may make RNs more likely to be complete and reduce the pressure to review changes:
\vspace{-1mm}
\begin{itemize}[label=\itembullet,leftmargin=10pt]
    \item For each change description in RNs, add links to the corresponding PR, issue, commit, or external resources (\eg CVE) so that its completeness can be easily checked.
    \item Adopt a systematic and structured way to label and organize changes (i.e., PRs/commits/issues), as discussed in Section~\ref{ss:regulation}. 
    \item Distribute workload among all contributors instead of having a central responsible person for creating RNs.
    For example, some projects require that each PR should contain a release notes section in the PR body that describes the affected submodule name and a list of changes for that submodule~\cite{Writing-Release-Notes:online}.
\end{itemize}
\vspace{-1mm}

\faCheckSquareO\;\textit{How Should RNs be Organized?}
The issues related to \textit{Presentation} indicate that layout indeed greatly influences RN reading experience, as mentioned by Bi et al.~\cite{bi2020empirical}.
An analogy is the relationship between content and directory: if the content is misplaced or not indexed, it is easy to miss the content you are interested in~\cite{electron_28375_issue:online,eos_9903_issue:online}. 
From these issues and their related RNs, we find that several hierarchical structures can be used to separate changes into categories and better organize RNs. 
Based on results in Section~\ref{sec:usablity}, we recommend two strategies to group changes: 1) by type of change (\eg new features, fixed bugs, breaking changes); 2) by affected component (\eg the network module).
The two strategies can be combined (e.g., first by component and then by type of change).
We also recommend putting the most important changes (e.g., breaking changes, major new features) on top.
After an organization is determined, we further recommend using proper visualization and fold lengthy lists for highlighting important changes.


\faCheckSquareO\;\textit{How to Choose Writing Style for RNs?}
When investigating issues under \textit{Bad Writing Style}, a case attracts our attention, \ie \textit{RN should be funny and cryptic in app stores to attract non-technical end users but concise and clear on GitHub to deliver information efficiently}.
Because the requirements of users differ from these of internal developers, we recommend projects to provide different RNs in different writing styles to serve different audiences (stakeholders).
For example, Apache Camel provides two types of RNs: one is more generalized and summarized~\cite{Camel311-Whatsnew:online}
intended for the end users, while the other is a list of all issues that have been resolved under this update intended for someone who needs technical details~\cite{camel.apache.org:online}.

%

\faCheckSquareO\;\textit{How to Make RNs (More) Accessible?}
This problem involves not only how users can access RN quickly (\textit{\textsf{Limited Exposure}} and \textit{\textsf{Lack Notification}}), but also where producers should collaboratively edit and store RNs (\textit{Where to Produce}).
It can be relieved through some more diverse ways for notification and access.
As summarized in Section~\ref{sec:accessibility}, we recommend developers to consider publicizing RNs in the following locations, if applicable, to make their RNs more accessible: GitHub Release Pages, Project Websites, Files in Repositories, Apps, and Instant Messaging Channels. 

\faCheckSquareO\;\textit{Link Check.}
We find many issues related to links, \eg \textit{Missing Links} and \textit{Wrong Links} under \textbf{Content}, and \textit{\textsf{Wrong/Broken Links to RNs}} under \textbf{Accessibility}.
Broken or wrong links often make users unpleasant and increase their cost of searching.
A developer from \Code{mantidproject/mantid} mentions that they need to go over release and check links work before releasing~\cite{mantid_31371_issue:online}. 
Checking invalid links regularly in RN can mitigate this problem, that can be achieved by some tools, \eg Xenu Link Sleuth~\cite{Xenus_link:online} and HTML Link Validator~\cite{HTMLLinkcheck:online}.
Besides, providing absolute path instead of relative path~\cite{hedgedoc_1114_pull:online}
can reduce potential broken risks, no matter in RNs or in other documents containing links to RNs, e.g., READMEs.

\vspace{-2mm}
\subsubsection{Automating RN Production}
\label{sec:automation}

Apart from the tool-specific problems in Section~\ref{sec:results-production}, we further summarize the following research directions that may greatly help automated RN generation:



\faGears\;\textit{Automated Labeling of Software Changes}:
Our results in Section~\ref{sec:results-automation} show that many developers request for tools to automate RN production.
However, to the best of our knowledge, existing tools have strong constraints on input.
Some well-known tools, \eg github-activity and Release Drafter, require a compatible PR label system.
Semantic-Release
requires developers to write commit messages following a specific rule, \ie Angular Commit Message Conventions, requiring developers to specify which category a commit belongs to manually.
These preconditions limit their application scope, and the whole project needs to change its production process to adapt to it~\cite{terraform-provider-opentelekomcloud_1164_pull:online}. 
Techniques for automated commit/PR classification, 
which we consider as a promising direction, can alleviate this problem. Existing commit classification methods (e.g.,~\cite{GHADHAB2021106566, 10.1145/3127005.3127016}) mainly focus on classifying commits into three maintenance categories (i.e., corrective, adaptive, and perfective) proposed by Swanson~\cite{10.5555/800253.807723}, which is not suitable for RN generation. Therefore, classifying commits into categories suitable for RN generation (e.g., the eight kinds of \textit{changes} proposed in Section~\ref{ss:checklist}) is needed to facilitate automated RN generation. 
Similar discrepancies also exist for works on PR classification~\cite{8422103, JIANG2021106394}.



\faGears\;\textit{Automated Summarization and Language Style Unification}:
As reflected in Section~\ref{sec:results-presentation}, a fluent and unified writing style is vital to RN \textit{\textsf{Readability}}.
However, existing tools generate RNs by integrating existing text, \eg PR titles and commit messages, which not only violates RNs' fundamental principle (\textit{it should focus on the impact for the user and make that
understandable}~\cite{Openstack_management:online}), but also offloads the quality responsibility to developers writing other development text. 
This often leads to poor readability of the final generated RNs.
With advances in natural language preprocessing (NLP) tasks like text summarization~\cite{el2021automatic} and style transfer~\cite{9551764}, it will be interesting to explore approaches that summarize existing development text and unify language style for automated RN generation. 




\faGears\;\textit{Automated Testing of RNs}: 
As shown in Section~\ref{sec:content}, \textit{\textsf{Completeness}} and \textit{\textsf{Correctness}} are the key to a high quality RN.
Although we synthesize a checklist of practices during the process of RN production, these largely manual practices are hardly a strong guarantee for reducing the risk of incompleteness or incorrectness.
To the best of our knowledge, there is still no tool designed for \textit{testing} (i.e., inconsistency checking) of RNs.
Challenges for facilitating such testing may include: 1) checking the consistency between natural language description and software changes; and 2) checking the consistency between documentation from different sources (e.g., RNs and usage guides, Section~\ref{sec:regulation}).
Similar works for, e.g., checking code comment inconsistency~\cite{tan2007icomment, zhai2020c2s}, may be a good starting port for exploring the possibility of such a tool.
Furthermore, since users perceive breaking changes as important but frequently missing in RNs (Section~\ref{sec:completeness}), works on breaking change and update incompatibility detection~\cite{lam2020putting} should also be important.

\vspace{-1.5mm}
\subsection{Threats to Validity}
\label{sec:threat}

\subsubsection{Internal Validity}
Our taxonomy construction is based entirely on manual analysis, which may introduce subjectivity and labeling errors.
To mitigate these threats, we include two inspectors and one arbitrator into the process, all with rich development experience.
To ensure the quality of taxonomy, we conduct multiple iterative rounds to refine the taxonomy and incorporates feedback from real developers.
We also measure inter-rater reliability to ensure that the taxonomy is precisely defined and reproducible.


\vspace{-1mm}
\subsubsection{External Validity}
Our work only uses issues from GitHub projects for categorizing RN issues, which means that our results may not be generalized to another context (e.g., industry projects).
Since GitHub is a huge and diverse coding platform and the projects involved in our analysis are of high quality, we believe our results reveal valuable insights and practical challenges in RN production and usage.
To further confirm our belief, we invite three industry developers to validate whether our taxonomy can cover the RN issues they have encountered.
However, the limited number of developers also poses a threat, which we find it hard to mitigate because it is not easy to locate industry developers experienced with RNs.
Future work may be able to gain different insights through other data sources or interviews/surveys on a larger scale.

Another threat to external validity comes from using only issues with keyword ``release note'' in their titles. 
Many issues may still discuss RNs even if they do not have the keyword in their titles.
The threat can be mitigated by the size of our dataset that is comparable to and even larger than existing studies~\cite{aghajani2019software,beyer2018automatically,zhang2019empirical,chen2020comprehensive, tan2019communicate}.

\vspace{-1.5mm}
\section{Conclusion}
\label{sec:conclusion}

In this paper, we have presented a taxonomy of real-world RN issues summarized from GitHub.
Our taxonomy not only distills a practitioner-oriented checklist for release note production, but also lays out an empirical foundation for several interesting research directions for release note automation.
As future work, we plan to investigate such opportunities for integrating novel automation approaches with existing release note workflows.

\vspace{2mm}
\noindent\textbf{Acknowledgments.} This work is supported by the National Key R\&D Program of China Grant 2018YFB1004201 and the National Natural Science Foundation of China Grant 61825201.

\bibliographystyle{ACM-Reference-Format}
\bibliography{thesis}

\end{document}